\documentclass[12pt]{article}

\usepackage[a4paper,left=30mm,right=20mm,top=20mm,bottom=20mm]{geometry}
\usepackage{graphics}
\usepackage{amsmath}
\usepackage{epsfig}
\usepackage{cite}

\title{Charmonia production from $b$-hadron decays at LHC with $k_T$-factorization: $J/\psi$, $\psi(2S)$ and $J/\psi + Z$} 
\author{A.V.~Lipatov$^{1,2}$, S.P.~Baranov$^{3}$, H.~Jung$^{4}$, M.A.~Malyshev$^{1}$}

\begin{document}

\maketitle

\begin{center}
{\it $^1$Skobeltsyn Institute of Nuclear Physics, Moscow State University, 119991 Moscow, Russia}\\
{\it $^2$Joint Institute for Nuclear Research, Dubna 141980, Moscow region, Russia}
{\it $^3$P.N.~Lebedev Physics Institute, 119991 Moscow, Russia}\\
{\it $^4$Deutsches Elektronen-Synchrotron, Notkestrasse 85, Hamburg, Germany}\\
\end{center}

\vspace{0.5cm}

\begin{center}

{\bf Abstract }

\end{center} 

\indent

We consider the production of $J/\psi$ and $\psi(2S)$ mesons
originating from the decays of $b$-flavored hadrons at the LHC
using the $k_T$-factorization approach. Our analysis covers
both inclusive charmonia production and production of $J/\psi$ mesons
in association with $Z$ bosons.
We apply the transverse momentum dependent (or unintegrated) 
gluon density in a proton derived from Catani-Ciafaloni-Fiorani-Marchesini (CCFM)
evolution equation and adopt fragmentation functions based on the 
non-relativistic QCD factorization to 
describe the inclusive $b$-hadron decays into the different charmonium states.
Our predictions agree well with latest experimental data taken by the 
CMS, ATLAS and LHCb Collaborations at $\sqrt s = 7$, $8$ and $13$~TeV.
The contributions from double parton scattering to the associated non-prompt $J/\psi + Z$ 
production are estimated and found to be small.

\vspace{1.0cm}

\noindent
PACS number(s): 12.38.-t, 12.38.Bx, 13.25.Hw, 14.40.Pq

\newpage

\section{Introduction} \indent

A few years ago, the CMS, ATLAS and LHCb Collaborations reported 
the measurements of the inclusive production of $J/\psi$ and $\psi(2S)$
mesons originating from $b$-hadron decays (so called "non-prompt" production) at the 
LHC energies $\sqrt s = 7$ and $8$~TeV\cite{1,2,3,4}.
More recently, the LHCb Collaboration has measured\cite{5} the total and differential cross sections of 
the inclusive non-prompt $J/\psi$ production
at $\sqrt s = 13$~TeV. Additionally, first experimental data on the associated 
non-prompt $J/\psi$ meson and $Z$ boson production
have been presented by the ATLAS Collaboration at $\sqrt s = 8$~TeV\cite{6}. 
These measurements provide a new opportunity for studying $b$-flavor 
production in $pp$ collisions at the LHC.
A special point of interest 
because it
involves both weak and strong
interactions and therefore serves as a complex test of electroweak theory,
perturbative Quantum Chromodynamics (pQCD) and our knowledge of parton distributions in
a proton.
The fixed-order next-to-leading-log (FONLL) approach\cite{7}
and the general-mass variable-flavor-number scheme (GM-VFNS)\cite{8} are able to
describe the inclusive production of non-prompt $J/\psi$ and $\psi(2S)$ mesons
over the whole charmonia transverse momentum range. 
The role of the next-to-leading (NLO) pQCD corrections to the associated non-prompt $J/\psi + Z$
production and contributions from the double parton scattering (DPS) mechanism
were investigated very recently\cite{9} using the \textsc{MadGraph} tool\cite{10}.
The main goal of our present study is to obtain first predictions 
for non-prompt $J/\psi$ and $\psi(2S)$ production, both inclusively and in association with a $Z$ boson,
employing the $k_T$-factorization approach\cite{11,12} and 
to give a systematic self-consistent analysis of recent CMS, ATLAS and LHCb data\cite{1,2,3,4,5,6}.
The $k_T$-factorization approach has certain technical advantages in the 
ease of including higher-order radiative corrections that can be taken into account 
in the form of transverse momentum dependent (TMD) 
parton distributions\footnote{See reviews\cite{13} for more 
information.}.

The calculations involve two main ingredients: the cross sections of $b$-hadron production 
and the partial widths of their subsequent decays into the different charmonia states.
In our previous papers\cite{14,15}, we paid attention to the first of them,
describing the inclusive production of $b$-hadrons and production of $Z$ bosons in association with
one or two $b$-jets (or rather $b$-hadrons) within the $k_T$-factorization formalism.
Our consideration below continues this line
and extends to the second ingredient. In addition to the previously tested
components of the theory, here we further inspect the evolution details of
$b$-quarks fragmenting and decaying into the final state charmonia.

The outline of our paper is following. In Section~2 we briefly describe 
our theoretical input and calculation steps. In Section 3 we present a
numerical results and a discussion. Section~4 contains our conclusions.

\section{The model} \indent

In the present note we strictly follow the approach described earlier\cite{14,15}.
Here we only briefly recall the main points of the theoretical scheme.

First, to calculate the cross sections of inclusive $b$-hadron production
and associated $Z + b$ production in $pp$ collisions the 
$k_T$-factorization approach was applied, 
mainly based on the off-shell gluon-gluon fusion subprocesses:
\begin{equation}
  g^* + g^* \to b + \bar b,
\end{equation}
\begin{equation}
  g^* + g^* \to Z + b + \bar b,
\end{equation}

\noindent
where the $Z$ boson further decays into a lepton pair.
The corresponding gauge-invariant off-shell (dependent on the transverse momenta 
of initial gluons) production amplitudes were calculated earlier (see \cite{14,15} and references therein)
and implemented into the Monte-Carlo event generator \textsc{cascade}\cite{16}.
Then, $b$-flavor production cross sections were calculated as a convolution of the off-shell 
partonic cross sections and TMD gluon 
distributions in a proton.

In addition to the off-shell gluon-gluon fusion, 
several subprocesses involving quarks in the initial state
are taken into account. These are the flavor excitation, 
quark-antiquark annihilation and quark-gluon scattering subprocesses:
\begin{equation} \label{qQ-ZqQ}
  q + b \to Z + q + b,
\end{equation}
\begin{equation} \label{qq-ZQQ}
  q + \bar q \to Z + b + \bar b,
\end{equation}
\begin{equation} \label{qg-ZqQQ}
  q + g \to Z + q + b + \bar b.
\end{equation}

\noindent
with subsequent decays of $Z$ bosons into a a lepton pair.
These processes become important at large transverse momenta $p_T$
(or, respectively, at large parton longitudinal momentum fraction $x$, which is needed to
produce high $p_T$ events) where the quarks are less suppressed or can even dominate 
over the gluon density. We find it reasonable to rely upon the collinear 
Dokshitzer-Gribov-Lipatov-Altarelli-Parisi (DGLAP) factorization scheme\cite{17}, 
which provides better theoretical grounds in the large-$x$ region.
We consider a combination of two techniques where each of them being used at the 
kinematic conditions where it is best suitable (off-shell gluon-gluon fusion 
subprocesses at small $x$ and quark-induced subprocesses at large $x$ values).

An essential point of our consideration is using a numerical solution of the 
Ciafaloni-Catani-Fiorani-Marchesini (CCFM) evolution
equation\cite{18} to derive the TMD gluon density in a proton. The CCFM equation 
provides a suitable tool since it smoothly interpolates 
between the small-$x$ Balitsky-Fadin-Kuraev-Lipatov (BFKL)\cite{19} gluon dynamics and high-$x$ DGLAP dynamics. 
Following\cite{15}, we adopt the latest JH'2013 parametrization\cite{20}, taking JH'2013 set~2 as the default choice.
This TMD gluon density was fitted to high-precision DIS data on the 
proton structure functions $F_2(x,Q^2)$ and $F_2^c(x,Q^2)$. 
For the conventional quark and gluon densities we use the MSTW'2008 (LO) set\cite{21}.
The fragmentation of the produced $b$ quarks into $b$-hadrons is described with the
Peterson fragmentation function with $\epsilon_b = 0.0126$\cite{22}.
Using these parameters, we successfully reproduce the latest CMS experimental 
data on inclusive $B^+$ meson production\footnote{In our previos studies\cite{14} older 
version of CCFM-evolved gluon density in a proton (namely, set A0\cite{23}) was 
applied to evaluate the inclusive $B^+$ meson production at the LHC.} at $\sqrt s = 13$~TeV\cite{24}.

The approach above provides the necessary starting point for the theoretical description
of $J/\psi$ or $\psi(2S)$ production from $b$-hadrons.
At the next step, the obtained $b$-hadron cross sections have to be convoluted with
$b\to J/\psi$ and $b\to\psi(2S)$ fragmentation functions. 
These fragmentation functions are the longitudinal momentum distributions
of the $J/\psi$ and $\psi(2S)$ mesons from $b$-hadron decay,
appropriately boosted along the $b$-hadron flight direction.
The latter have been calculated\cite{25} in the framework of 
nonrelativistic QCD (NRQCD) factorization\cite{26,27} using the approach\cite{28}.
In this approach, the decays of $b$-hadrons into the final charmonia are
represented by a sum of products. Each of them consists of a
perturbative coefficient for the production of a $c\bar c$ pair 
in a specific angular-momentum and color state and 
nonperturbative NRQCD matrix element (NME). It 
parametrizes the subsequent transition of this intermediate $c \bar c$ state 
into the final physical charmonium with 
the help of nonperturbative soft gluon radiation
and it has to be extracted from the experimental data.
The resulting predictions\cite{25} for the $b \to J/\psi + X$ and 
$b \to \psi(2S) + X$ three-momentum distributions 
reasonably agree with the measurements performed
by the CLEO\cite{29} and BABAR\cite{30} Collaborations.
Then, the developed formalism\cite{25} is implemented into our calculations
without any changes\footnote{In particular, we took the NMEs from\cite{25}.
It is well-known that these NMEs depend on the minimal charmonia transverse momentum
used in the fits and are incompatible with each other 
when obtained from fitting the
different sets of data.
However, it was argued\cite{25} that the fine details
of the developed formalism, like as exact values of the NMEs, are
almost irrelevant (at least at large $p_T$) due to the Lorentz
boost from the $b$-hadron rest frame to the laboratory one, so that 
the corresponding branching fraction becomes the key parameter.}.
To be precise, we employ the asymptotic expression\cite{25} for the $b$-hadron decay
distribution differential in the longitudinal momentum fraction $z$
carried by the produced charmonium state, obtained in the limit $|{\mathbf p}_b| \gg m_b$,
where $p_b$ and $m_b$ are the momentum and the mass of decaying $b$-hadron.
This approximation is valid within $11$\% and $5$\% accuracy 
for $|{\mathbf p}_b| = 10$~GeV and $20$~GeV, respectively, that is
suitable for our phenomenological study.

According to the experimental setup\cite{1,2,3,4,5,6}, we also included the
feed-down contributions to $J/\psi$ production from the
$b \to \chi_{cJ} + X$ with $J = 0, 1, 2$ and $b \to \psi(2S) + X$ decays
(followed by their subsequent radiative decays $\chi_{cJ} \to J/\psi + \gamma$ 
and $\psi(2S) \to J/\psi + \gamma$), taking them into account using the same 
approach\cite{25}.
The $\psi(2S)$ mesons are produced with no significant contributions
from decays of higher-mass quarkonia.
Following\cite{31}, we set the branching fractions $B(b \to J/\psi + X) = 0.68$\%,
$B(b \to \psi(2S) + X) = 0.18$\%, $B(b \to \chi_{c0} + X) = 0.015$\%,
$B(b \to \chi_{c1} + X) = 0.21$\%, $B(b \to \chi_{c2} + X) = 0.026$\%, 
$B(\psi(2S) \to J/\psi + \gamma) = 61$\%,
$B(\chi_{c0} \to J/\psi + \gamma) = 1.27$\%,
$B(\chi_{c1} \to J/\psi + \gamma) = 33.9$\%,
$B(\chi_{c2} \to J/\psi + \gamma) = 19.2$\%.
Numerically, we set $m_{J/\psi} = 3.097$~GeV, $m_{\psi(2S)} = 3.686$~GeV,
$m_{\chi_{c0}} = 3.415$~GeV, $m_{\chi_{c1}} = 3.511$~GeV, $m_{\chi_{c2}} = 3.556$~GeV\cite{31}.
Other essential parameters, such as renormalization and factorization scales, masses of produced particles 
are taken exactly the same as in our previous studies\cite{14,15}.

\section{Numerical results} \indent

Now we are in a position to present numerical results and a discussion.
We consider first the inclusive non-prompt $J/\psi$ and $\psi(2S)$ production.
The latest measurements have been carried out by the CMS\cite{1}, ATLAS\cite{2} 
and LHCb\cite{3,4,5} Collaborations.
The CMS Collaboration presented non-prompt $J/\psi$ transverse momentum $p_T$
distributions at $\sqrt s = 7$~TeV for three subdivisions in 
$J/\psi$ rapidity: $|y| < 1.2$, $1.2 < |y| < 1.6$ and $1.6 < |y| < 2.4$. The ATLAS Collaboration
measured both $J/\psi$ and $\psi(2S)$ transverse momentum distributions at $\sqrt s = 7$ and $8$~TeV
for eight rapidity subdivisions. The LHCb Collaboration presented the data on the 
inclusive non-prompt $J/\psi$ production
in the range $p_T < 14$~GeV and $2 < y < 4.5$ at different energies $\sqrt s = 7$, $8$ and $13$~TeV.
The $J/\psi$ and $\psi(2S)$ cross sections measured by the CMS and ATLAS Collaborations
were multiplied by the corresponding branching fractions $B(J/\psi \to \mu^+ \mu^-) = 5.961$\% and
$B(\psi(2S) \to \mu^+ \mu^-) = 0.79$\%\cite{31}, respectively.

We confront our predictions with the available data in Figs.~1 --- 4.
The solid histograms represent our central predictions
by fixing both the renormalization $\mu_R$ and factorization $\mu_F$ scales at their
default values (see\cite{14,15} for the detailed description of our input), while the shaded regions
correspond to the scale uncertainties of our predictions.
To estimate the latter we used the JH'2013 set 2$+$ and 
JH'2013 set 2$-$ sets instead of default one JH'2013 set 2. These two sets 
represent a variation of the renormalization scale used in the 
off-shell production amplitude\cite{20}.
We achieve good agreement with the LHC data on the 
transverse momentum distributions for both $J/\psi$ and $\psi(2S)$ mesons
within the experimental and theoretical uncertainties.
The slight disagreement is only observed at low 
transverse momenta, $p_T \le 10$~GeV. It could be 
attributed to the finite $|{\mathbf p}_b|$ terms in the 
longitudinal momentum distributions of the charmonia (important 
at $|{\mathbf p}_b| \sim m_b$), which are not taken into account in our calculations.
The rapidity distributions of $J/\psi$ mesons measured by the LHCb Collaboration
at different energies are well reproduced.
Our results are consistently close to 
the FONLL\cite{7} predictions presented in\cite{1,2,3,4,5},
which is explained by the fact that the main part of collinear QCD higher-order corrections
(namely, NLO + NNLO + N$^3$LO + ... contributions which correspond to the $\log 1/x$ enhanced 
terms in the perturbative series) are effectively taken into account as a part of the CCFM gluon 
evolution. Here we demonstrate again the main advantage of the $k_T$-factorization approach,
which gives us the possibility to estimate the size of these higher-order corrections
and reproduce in a straighforward manner the main features
of rather cumbersome higher-order pQCD calculations.

The estimated non-prompt $J/\psi$ production fiducial cross sections at $\sqrt s = 7$, $8$ and $13$~TeV 
are listed in Table~1 in comparison with the available LHC data.
One can see that our predictions agree well with these 
data within the theoretical and experimental uncertainties.

\begin{table}
\begin{center}
\begin{tabular}{|c|c|c|}
\hline
  & & \\
   Source  & $\sigma(J/\psi)$, predicted & $\sigma(J/\psi)$, measured \\
  & & \\
\hline
  & & \\
  CMS\cite{1}, $7$~TeV [nb] & $24.75^{+2.90}_{-1.07}$ & $26.0\pm 1.4$(stat.)$\pm 1.6$(syst.)$\pm 2.9$(lumi.) \\
  & & \\
  LHCb\cite{3}, $7$~TeV [$\mu$b] & $1.20^{+0.26}_{-0.11}$ & $1.14\pm 0.01$(stat.)$\pm 0.16$(syst.) \\
  & & \\
  LHCb\cite{4}, $8$~TeV [$\mu$b] & $1.39^{+0.29}_{-0.12}$ & $1.28\pm 0.01$(stat.)$\pm 0.11$(syst.) \\
  & & \\
  LHCb\cite{5}, $13$~TeV [$\mu$b] & $2.28^{+0.39}_{-0.16}$ & $2.25\pm 0.01$(stat.)$\pm 0.14$(syst.) \\
  & & \\
\hline
\end{tabular}
\caption{The fiducial cross sections of inclusive non-prompt $J/\psi$
production at $\sqrt s = 7$, $8$ and $13$~TeV. 
The experimental data are from CMS\cite{1} and LHCb\cite{3,4,5}. 
The cross section reported by the CMS Collaboration has been multiplied by the $J/\psi \to \mu^+ \mu^-$
branching fraction.}
\label{table1}
\end{center}
\end{table}

Now we turn to associated non-prompt $J/\psi + Z$ production.
First experimental data were obtained recently by the ATLAS Collaboration at $\sqrt s = 8$~TeV\cite{6}.
The leptons originating from the $Z$ boson decay 
are required to have pseudorapidities $|\eta^l| < 2.5$, transverse momenta 
$p_T^l > 25$~GeV (leading lepton) and $p_T^l > 15$~GeV (sub-leading lepton)
and invariant mass of the lepton pair $M^{ll}$ should lie within the interval
$|M^{ll} - m_Z| < 10$~GeV. The produced $J/\psi$ meson is required to have
the transverse momentum $8.5 < p_T < 100$~GeV and rapidity $|y| < 2.1$.
Fiducial cross sections were measured as ratios 
to the inclusive $Z$ boson production rate in the same fiducial volume.
We implemented this experimental setup into our numerical program.
The results of our calculations are shown in Fig.~5. One can see that,
similar to inclusive $J/\psi$ production case, 
our predictions for associated $J/\psi + Z$ production
reasonably agree with the ATLAS data\cite{6}.
We find that both the transverse momentum and rapidity distributions 
of the produced $J/\psi$ mesons are described within the theoretical and experimental uncertainties.
However, we note that the overall description could be even improved 
if additional contributions from the DPS mechanism are taken into
account.
To estimate the latter we apply a simple factorization
formula (for details see the reviews\cite{32,33,34} and references therein):
\begin{equation}
  \sigma_{\rm DPS} (J/\psi + Z) = {\sigma(J/\psi) \, \sigma (Z) \over \sigma_{\rm eff}}, 
\end{equation}

\noindent
where $\sigma_{\rm eff}$ is a normalization constant which incorporates all 
"DPS unknowns" into a single phenomenological parameter.
A numerical value of $\sigma_{\rm eff} \simeq 15$~mb was obtained from fits to $pp$ and $p\bar{p}$ 
data (see, for example,\cite{35}). 
We find that the DPS mechanism gives $\sim 10$\% contribution
to the production cross section 
in the considered kinematical region and populates mainly at low transverse momenta (see Fig.~5).
Being added to the predictions of usual single parton scattering (SPS) mechanism, 
the DPS contributions slightly improve the description of the 
$J/\psi$ rapidity distribution in a whole $y$ range. 
Some reasonable variations in $\sigma_{\rm eff} \simeq 15 \pm 5$~mb
would affect DPS predictions, though without changing our basic 
conclusion\footnote{The lower limit of $\sigma_{\rm eff} \sim 5$~mb was established\cite{9} using 
the \textsc{MadGraph} tool.}.
The presented DPS estimation coincides with the earlier one performed by the ATLAS Collaboration\cite{6}.

\section{Conclusions} \indent

We investigated the production of $J/\psi$ and $\psi(2S)$ mesons,
originating from the decays of $b$-flavored hadrons,
both inclusively and in association with $Z$ bosons,
at the LHC conditions using the $k_T$-factorization approach. 
Our consideration was mainly based 
on the dominant off-shell gluon-gluon fusion subprocesses where 
the transverse momenta of initial gluons are taken into account.
In the case of associated non-prompt $J/\psi + Z$ production, a number of subleading quark-induced 
subprocesses have been considered in the conventional collinear scheme.
To describe the inclusive $b$-hadron decays into the different charmonium states,
we applied fragmentation functions, based on the NRQCD factorization.
The feed-down contributions from $\chi_{cJ}$ (with $J = 0, 1, 2$) and 
$\psi(2S)$ decays to $J/\psi$ meson production were taken into account.

Using the TMD gluon densities derived from the CCFM gluon evolution 
equation, we achieved a good agreement between our 
predictions and latest CMS, ATLAS and LHCb data collected at $\sqrt s = 7$, $8$ and $13$~TeV. 
We provided first theoretical expectations for the associated 
non-prompt $J/\psi + Z$ production and estimated the contribution of 
the double parton scattering mechanism to the production cross sections.
The latter is found to be of order of $\sim 10$\%.

\section{Acknowledgements} \indent

We thank F.~Hautmann for very useful discussions
and remarks. 
This research was supported by RFBR grant 16-32-00176-mol-a and
grant of the President of Russian Federation NS-7989.2016.2.
We are grateful to DESY Directorate for the
support in the framework of Moscow --- DESY project on Monte-Carlo implementation for
HERA --- LHC. M.A.M. was also supported by a grant of the foundation for
the advancement of theoretical physics "Basis" 17-14-455-1.

\newpage 

\begin{figure}
\begin{center}
\includegraphics[width=7.9cm]{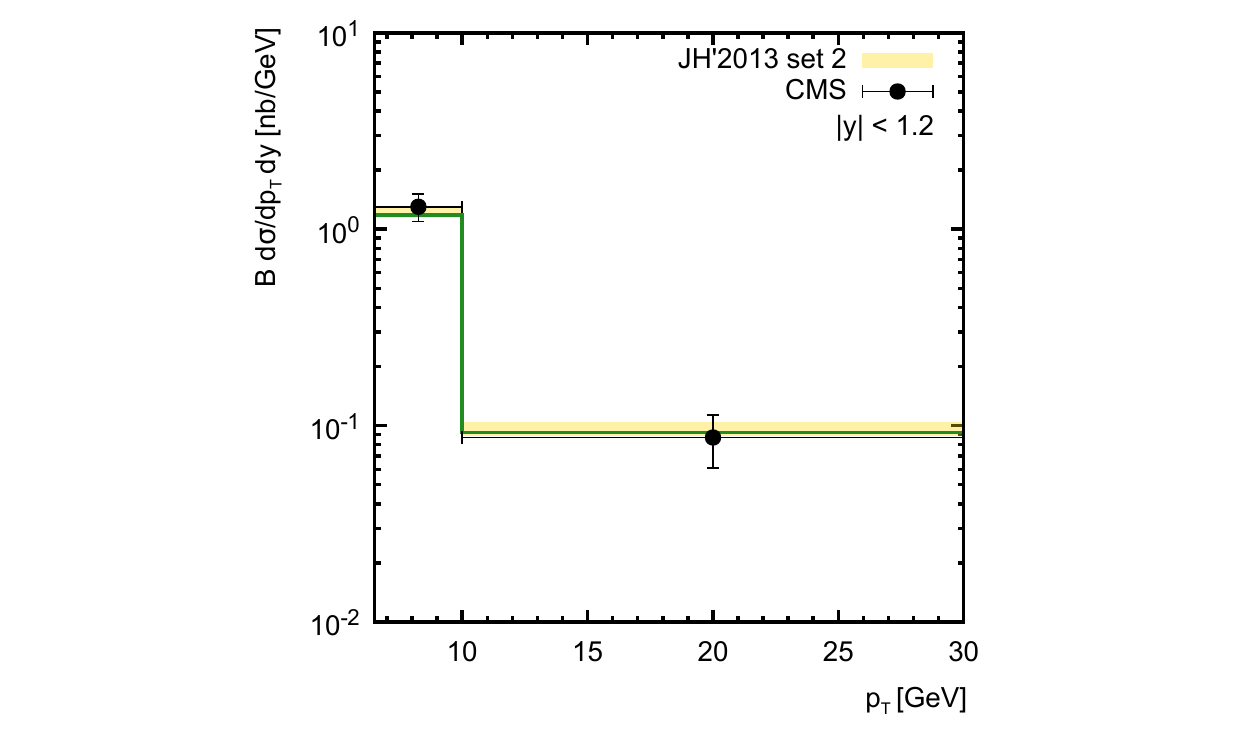}
\includegraphics[width=7.9cm]{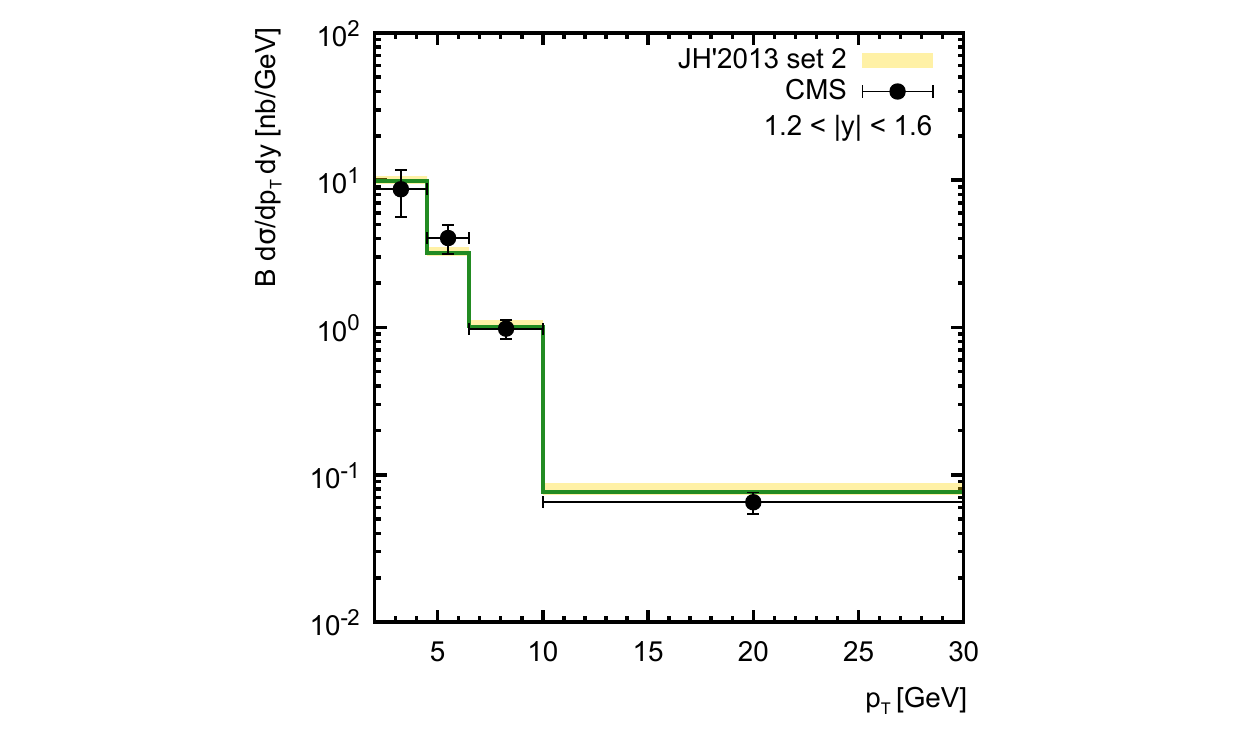}
\includegraphics[width=7.9cm]{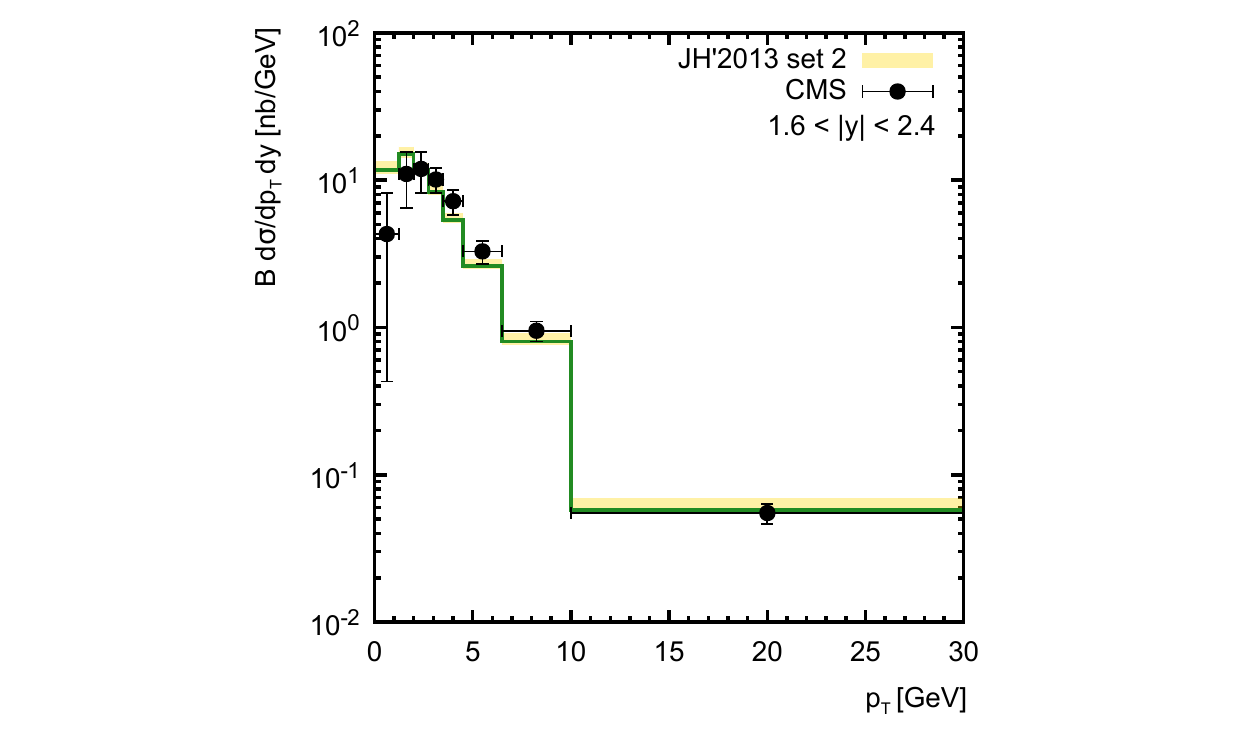}
\caption{The double differential cross sections of inclusive non-prompt $J/\psi$ meson
production at $\sqrt s = 7$~TeV as a function of $J/\psi$ transverse momentum. 
The solid histograms represent our predictions obtained with the JH'2013 set 2 gluon density
at the default hard scales. The shaded bands represent the scale uncertainties of the calculations, as it is 
described in the text. The experimental data are from CMS\cite{1}.}
\label{fig1}
\end{center}
\end{figure}

\begin{figure}
\begin{center}
\includegraphics[width=7.9cm]{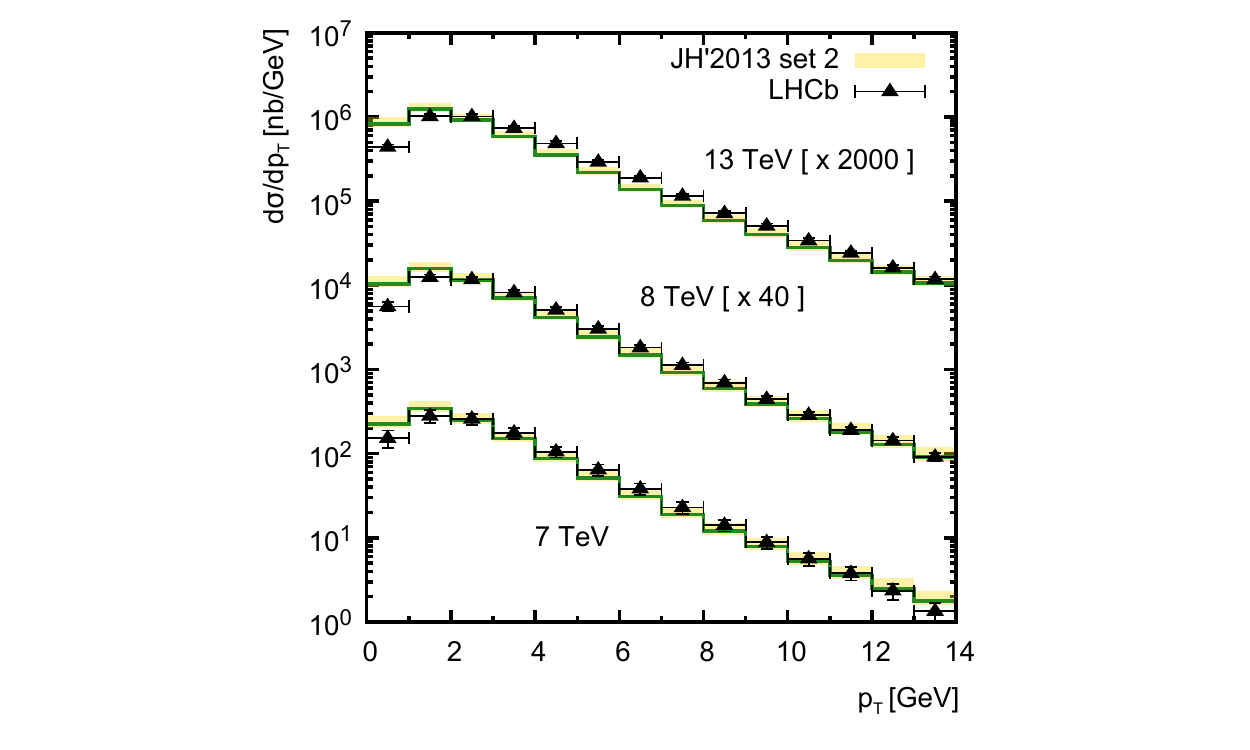}
\includegraphics[width=7.9cm]{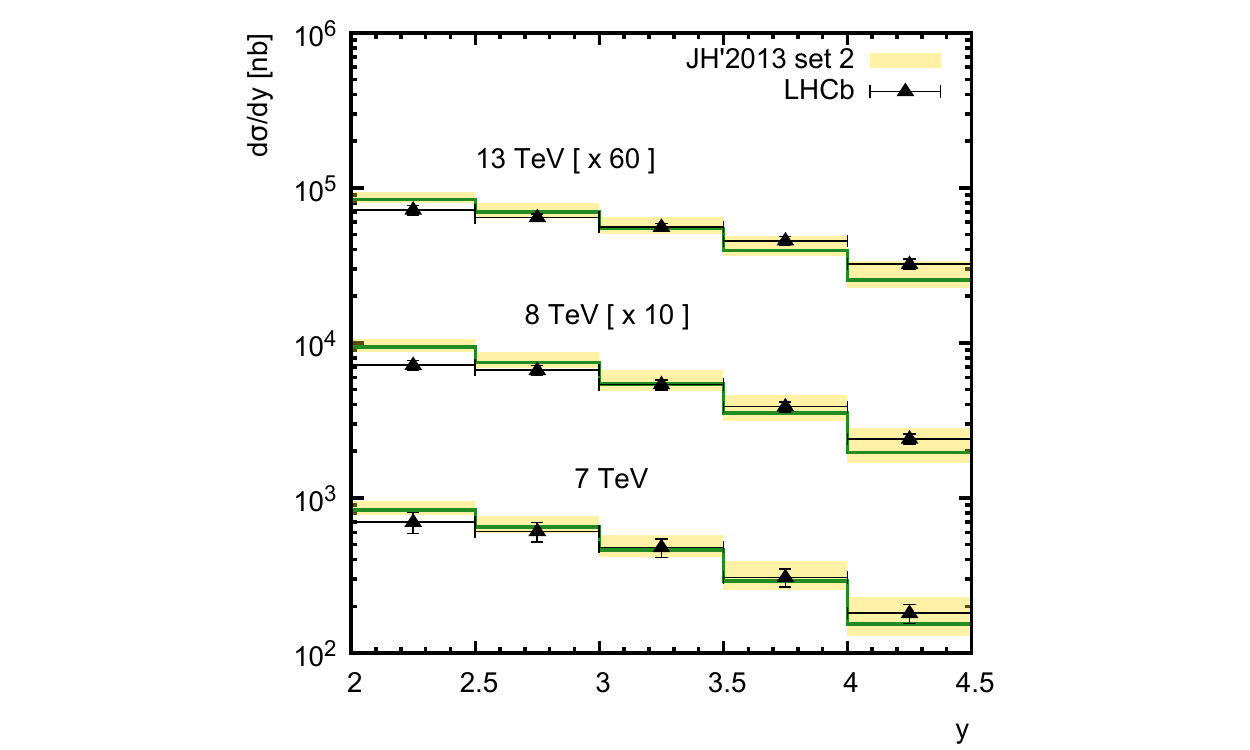}
\caption{The differential cross sections of inclusive non-prompt $J/\psi$ meson
production at $\sqrt s = 7$, $8$ and $13$~TeV as functions of $J/\psi$ transverse momentum and rapidity.
Notation of curves is the same as in Fig.~1. The experimental data are from LHCb\cite{3,4,5}.}
\label{fig2}
\end{center}
\end{figure}

\begin{figure}
\begin{center}
\includegraphics[width=7.9cm]{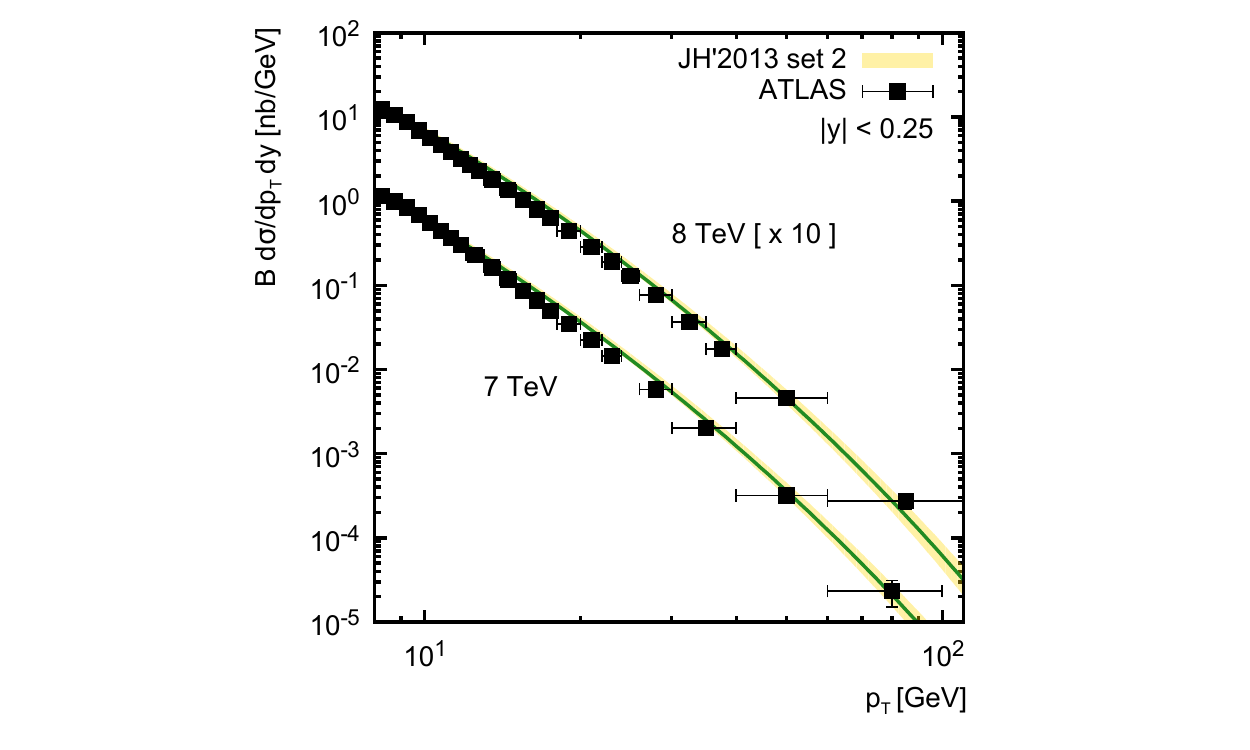}
\includegraphics[width=7.9cm]{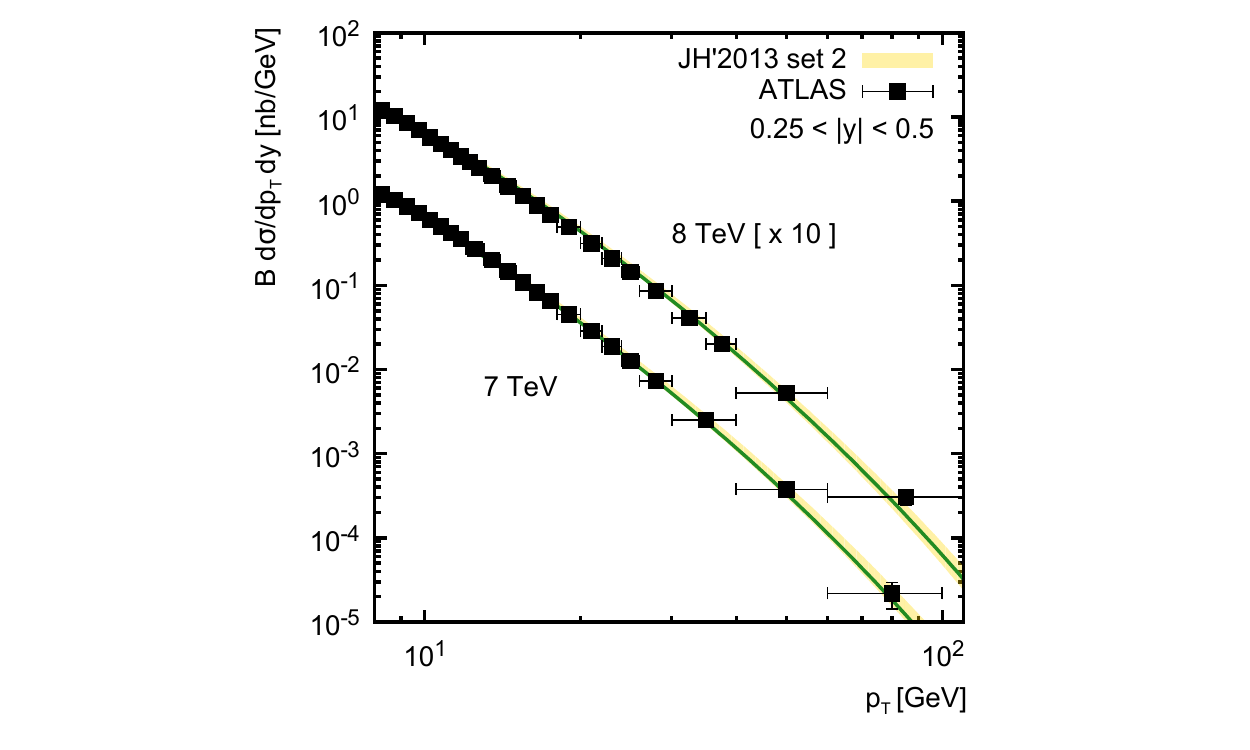}
\includegraphics[width=7.9cm]{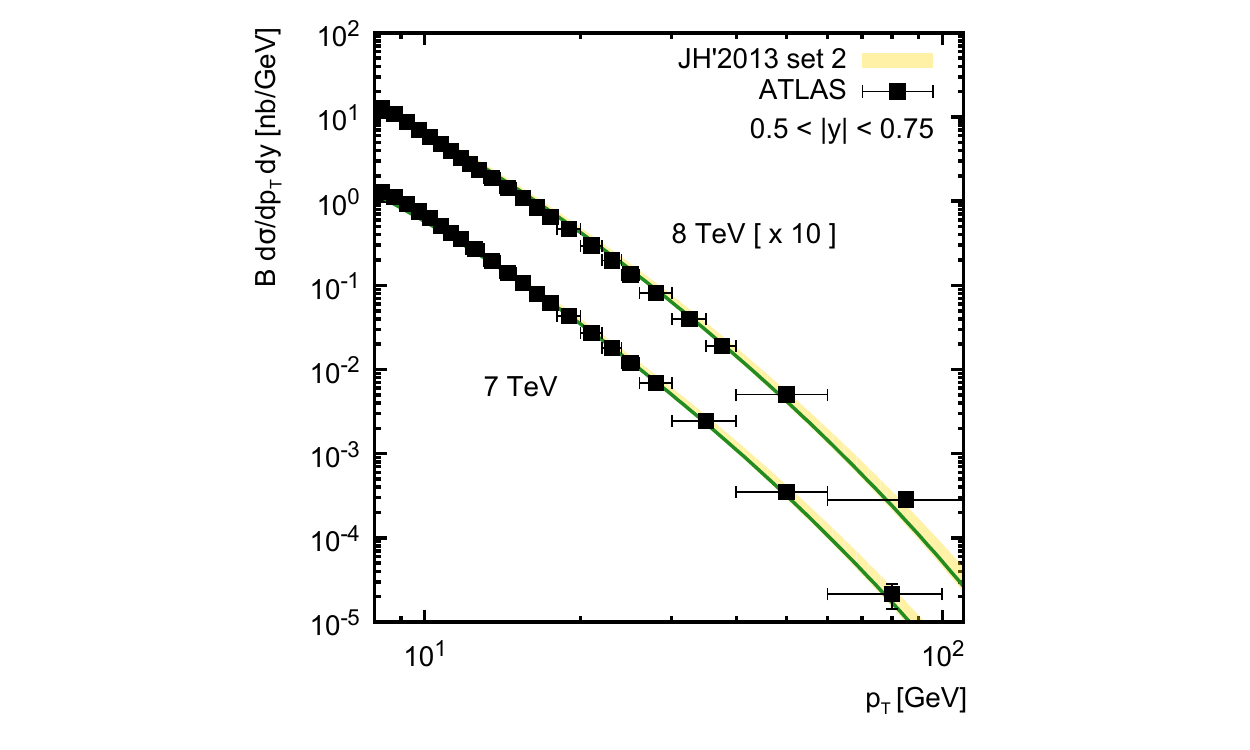}
\includegraphics[width=7.9cm]{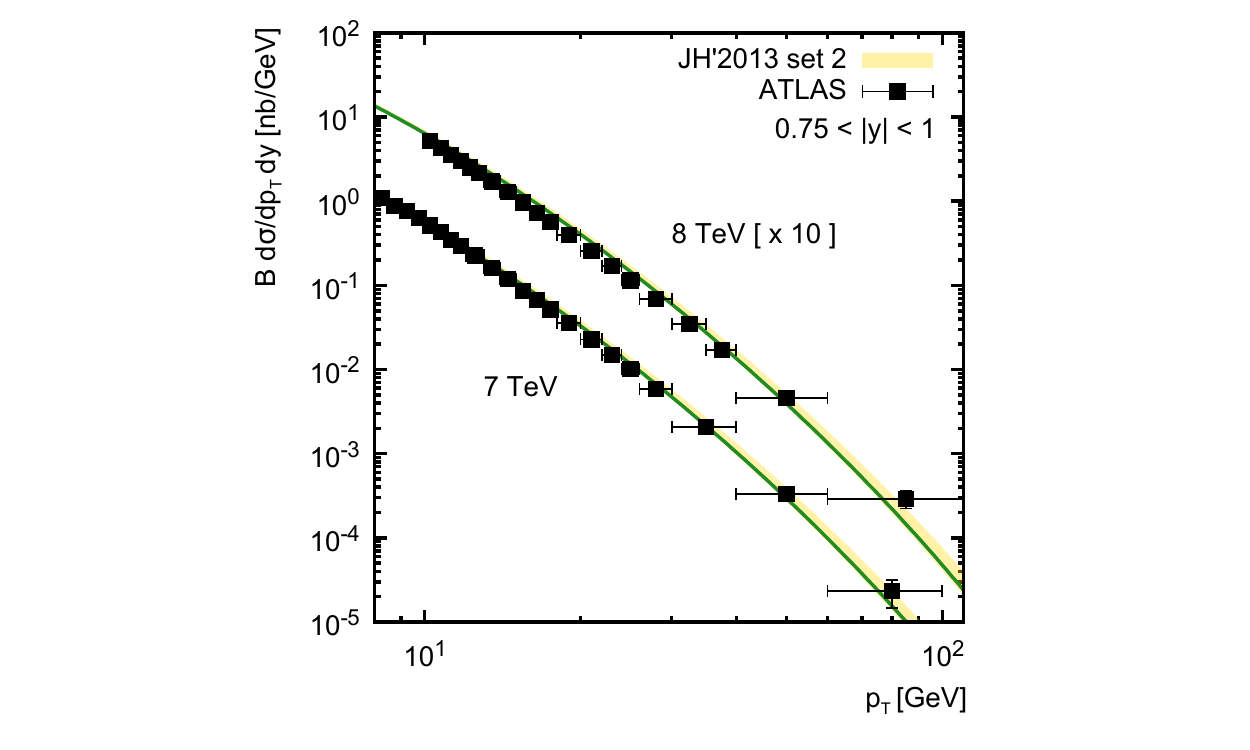}
\includegraphics[width=7.9cm]{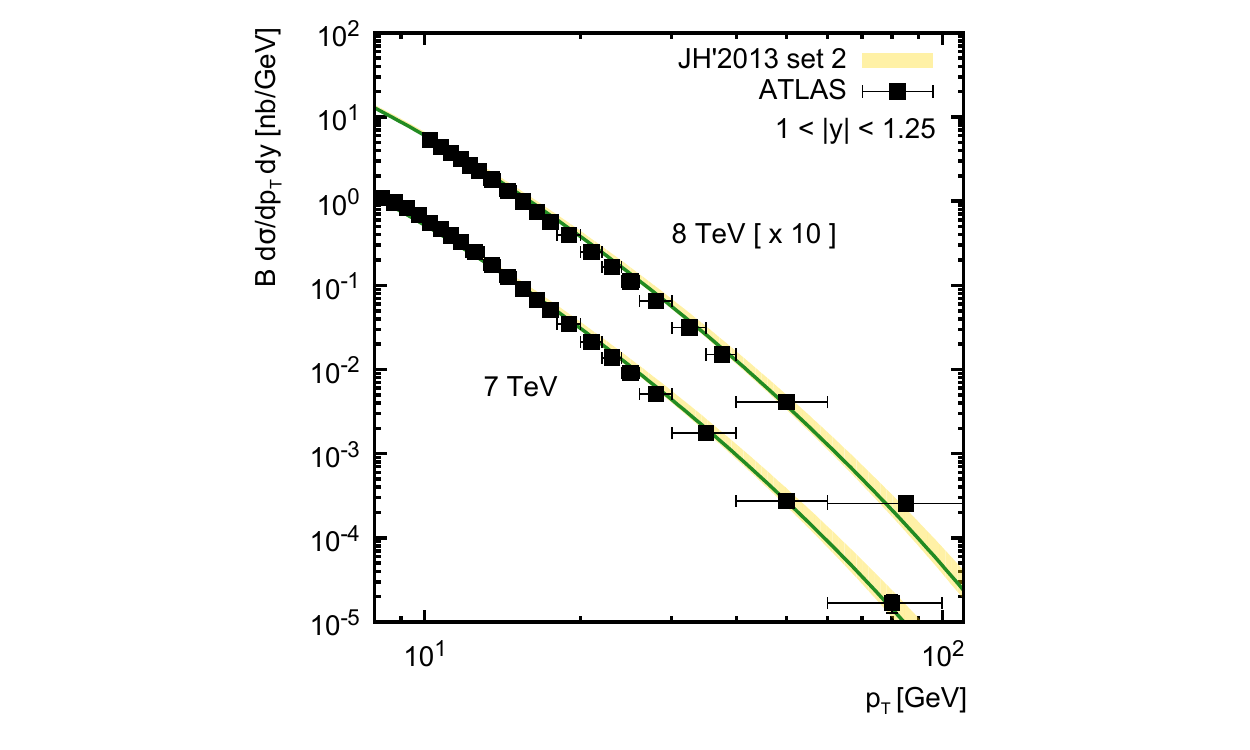}
\includegraphics[width=7.9cm]{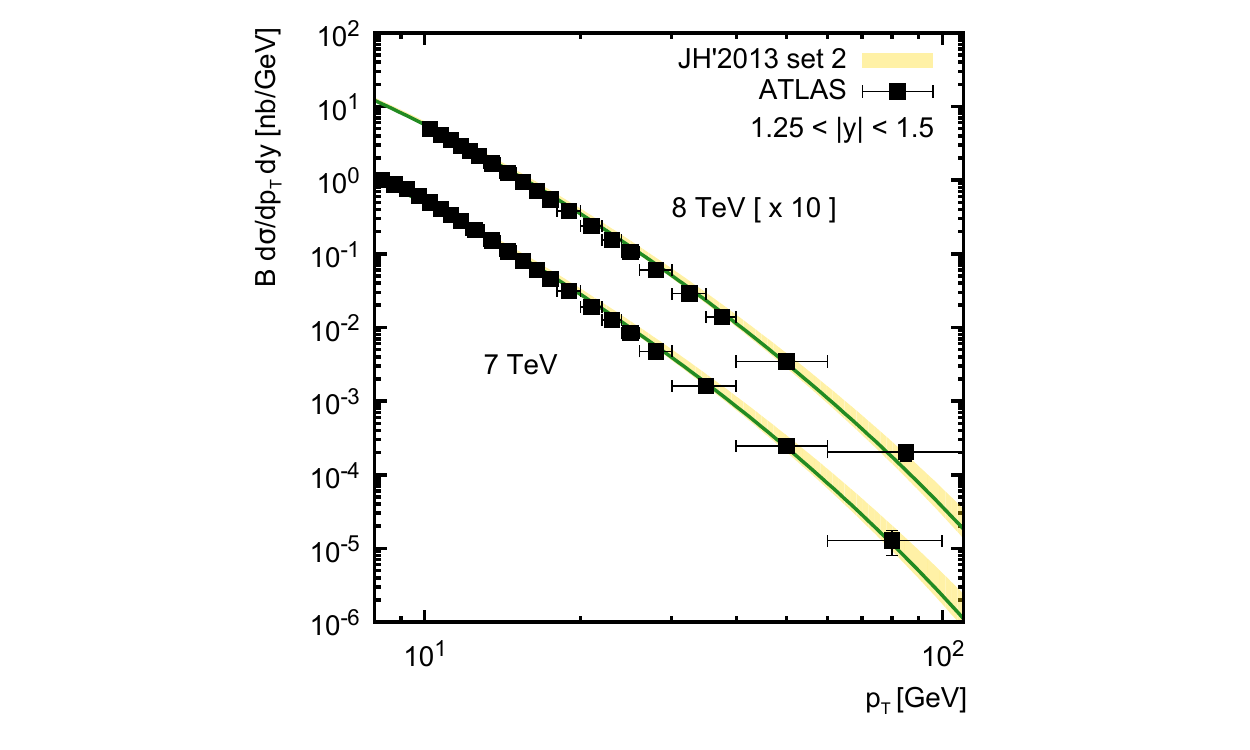}
\includegraphics[width=7.9cm]{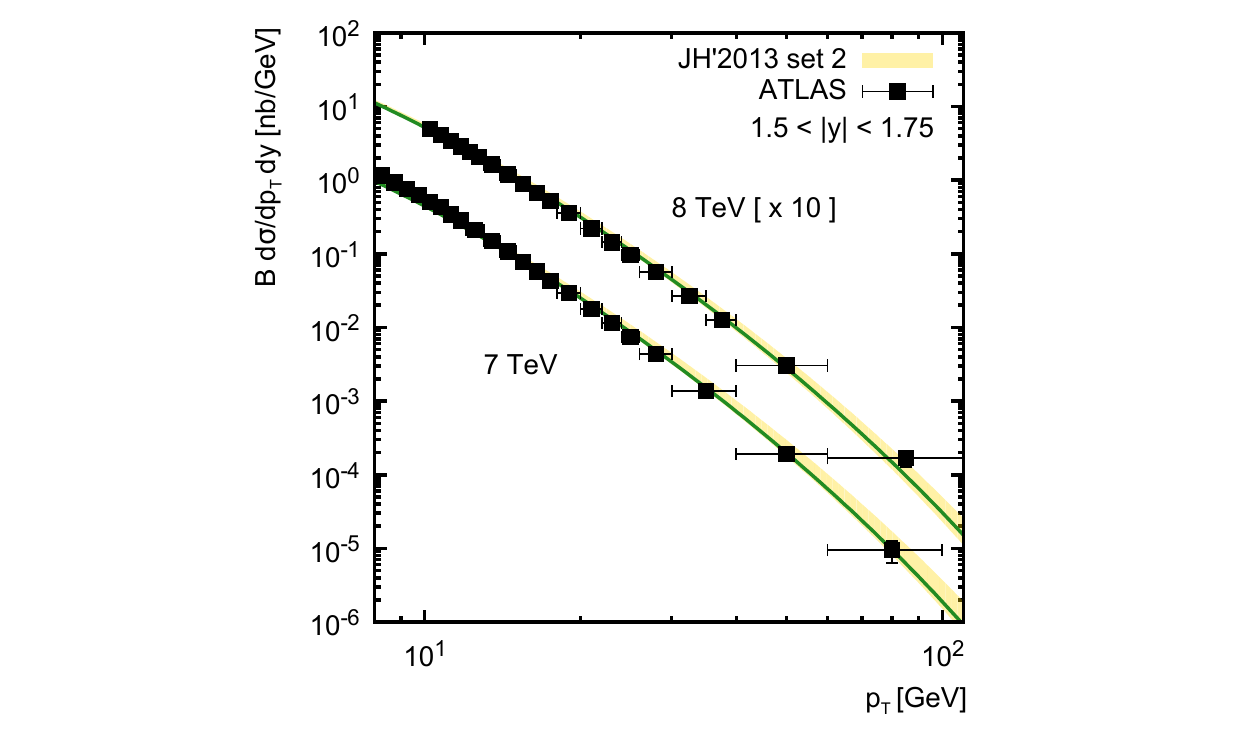}
\includegraphics[width=7.9cm]{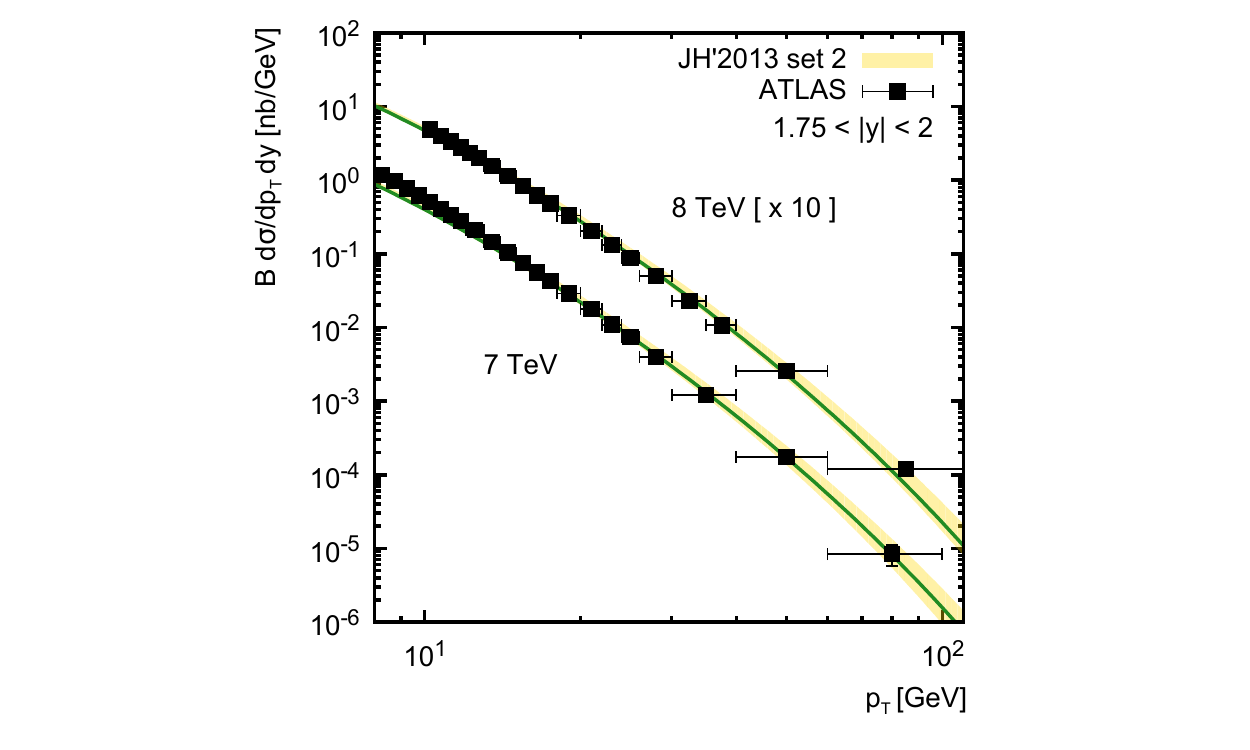}
\caption{The double differential cross sections of inclusive non-prompt $J/\psi$ meson
production at $\sqrt s = 7$ and $8$~TeV as a function of $J/\psi$ transverse momentum.
Notation of curves is the same as in Fig.~1. The experimental data are from ATLAS\cite{2}.}
\label{fig3}
\end{center}
\end{figure}

\begin{figure}
\begin{center}
\includegraphics[width=7.9cm]{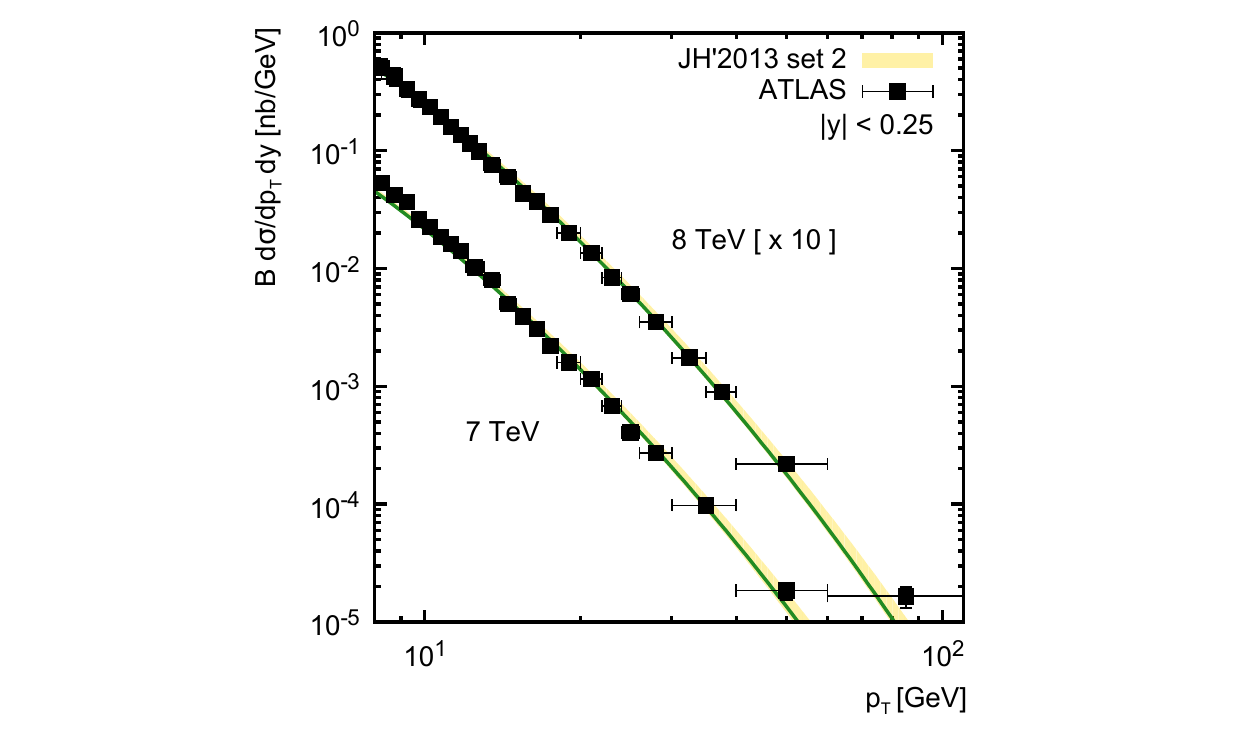}
\includegraphics[width=7.9cm]{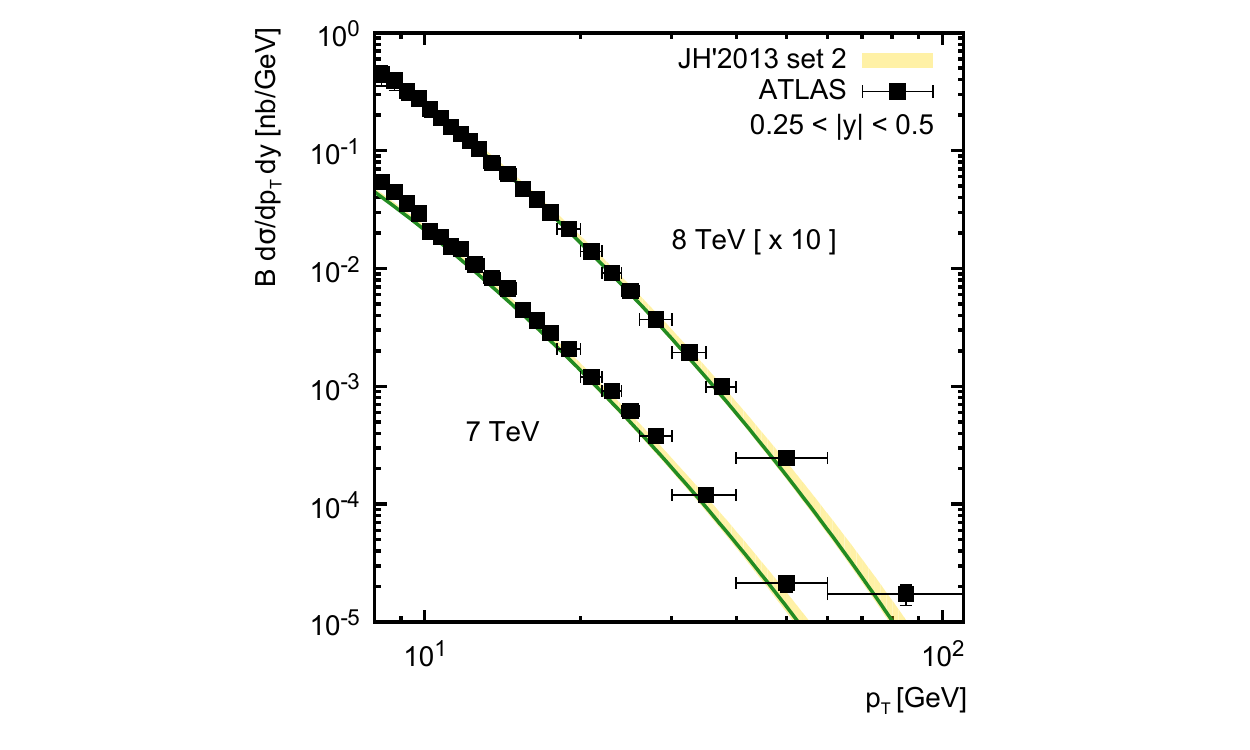}
\includegraphics[width=7.9cm]{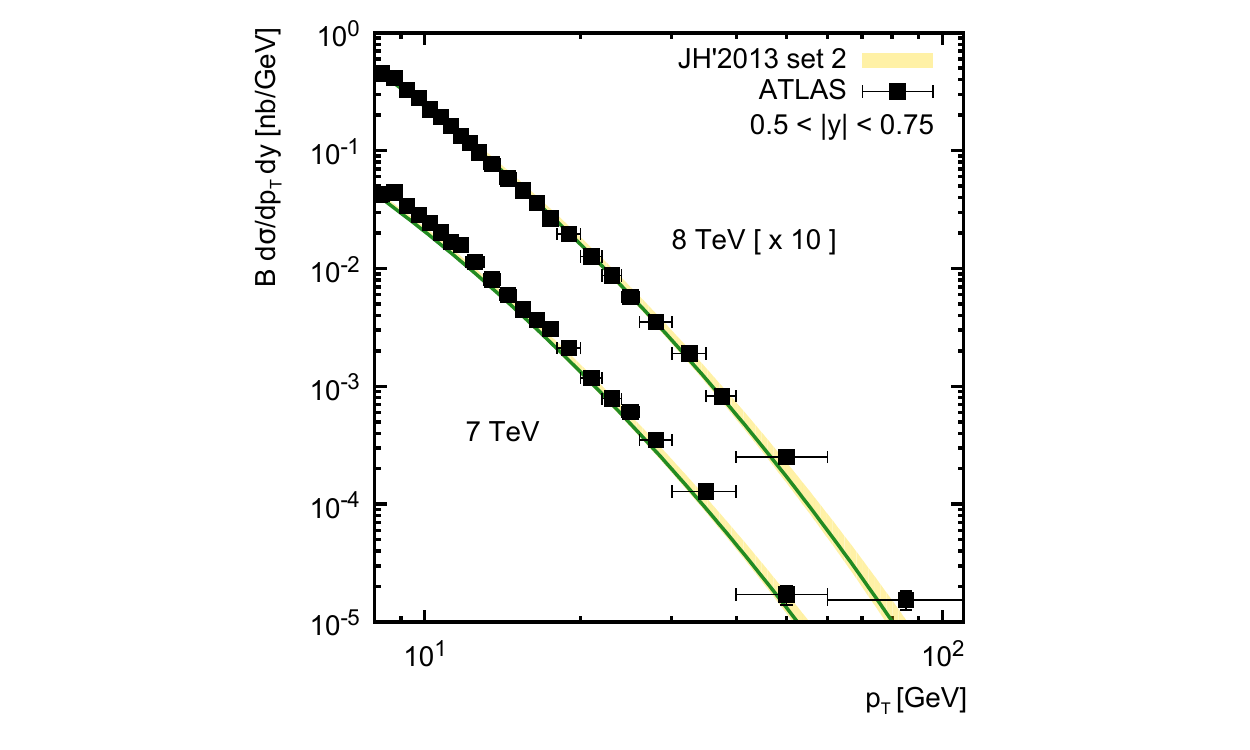}
\includegraphics[width=7.9cm]{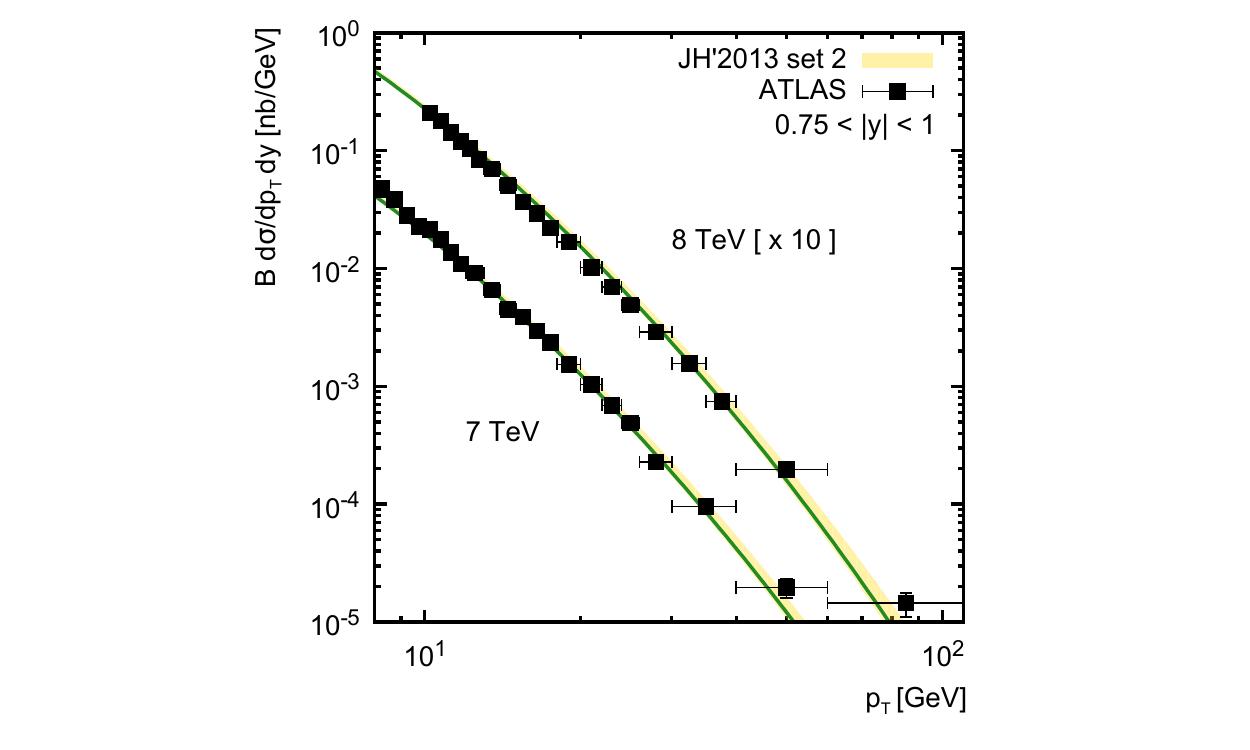}
\includegraphics[width=7.9cm]{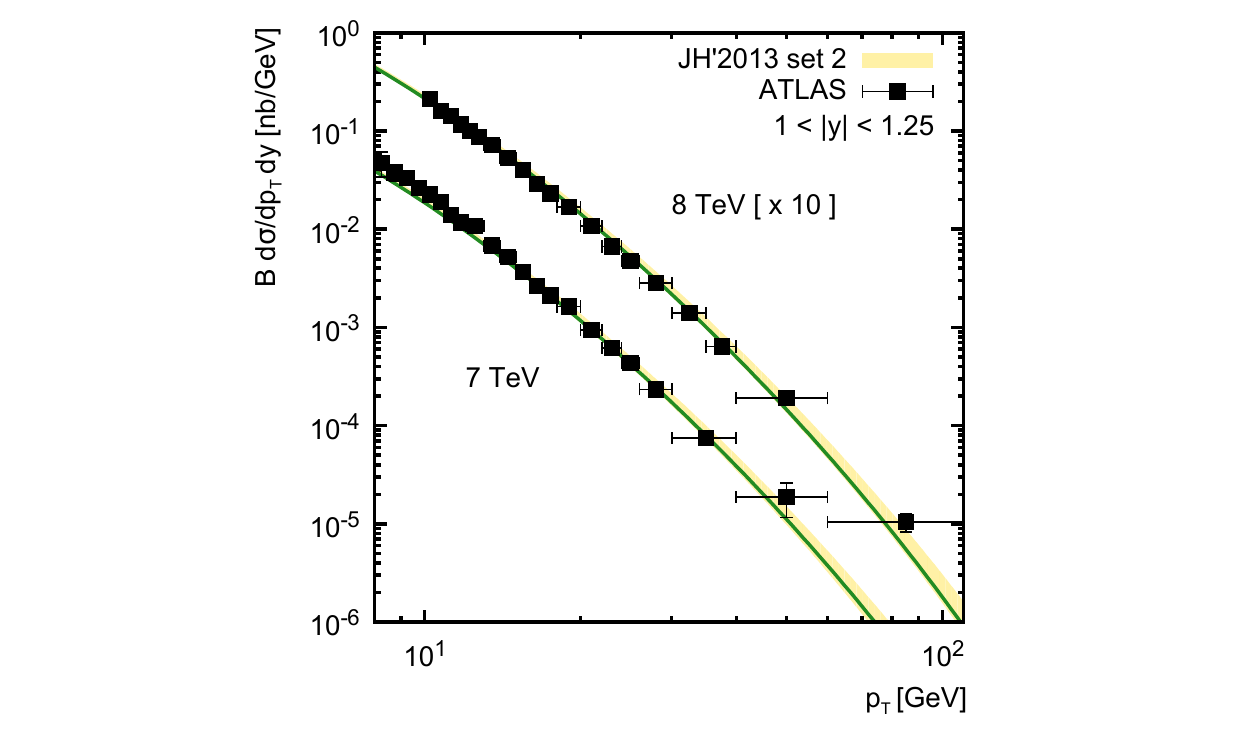}
\includegraphics[width=7.9cm]{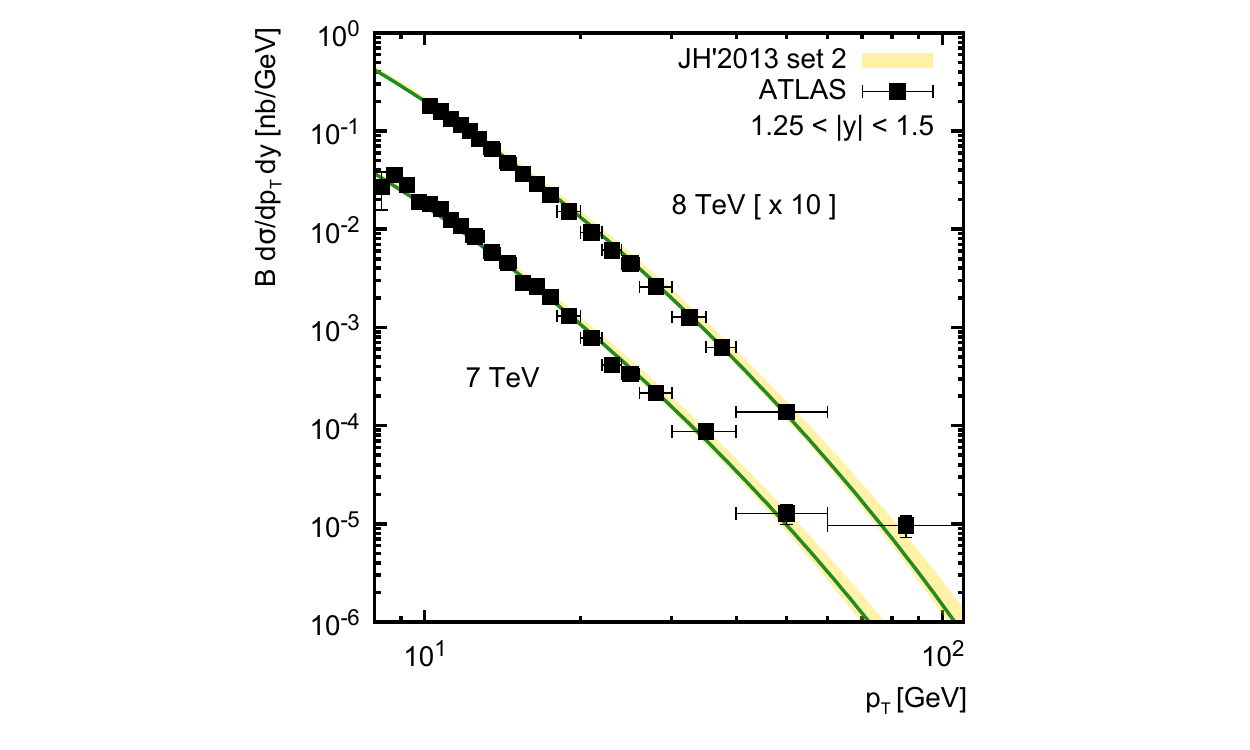}
\includegraphics[width=7.9cm]{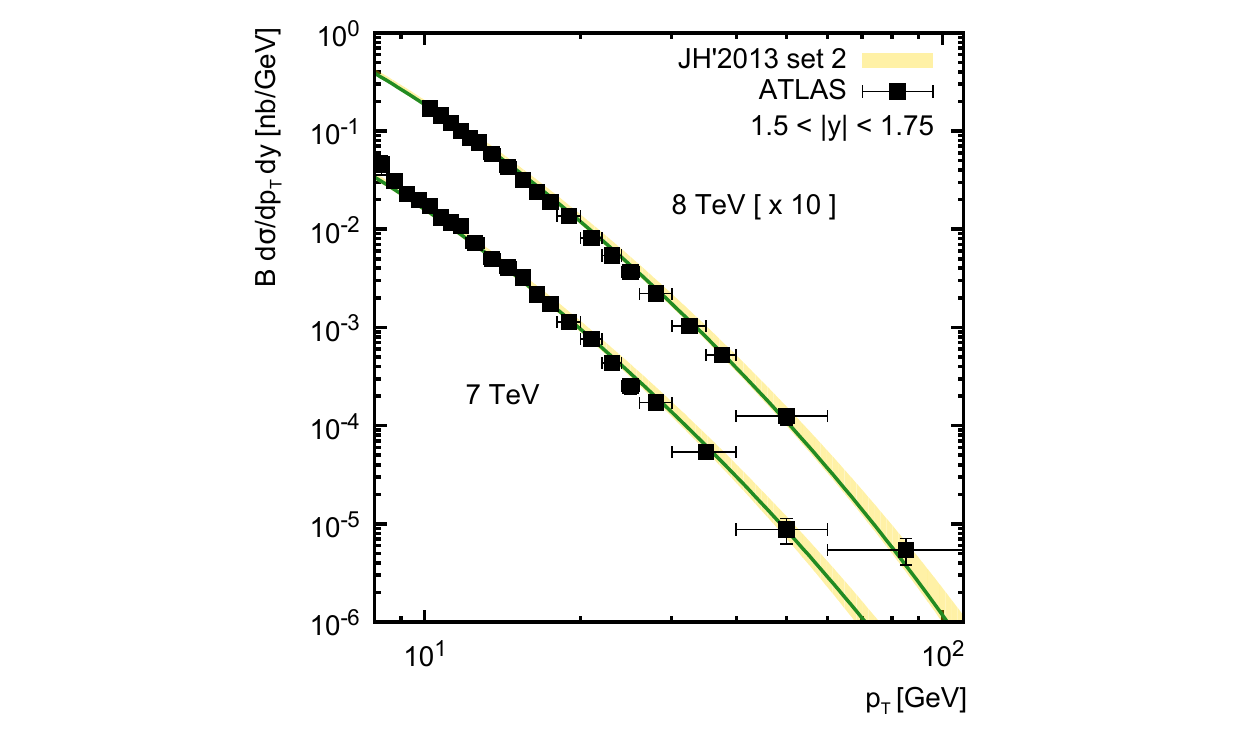}
\includegraphics[width=7.9cm]{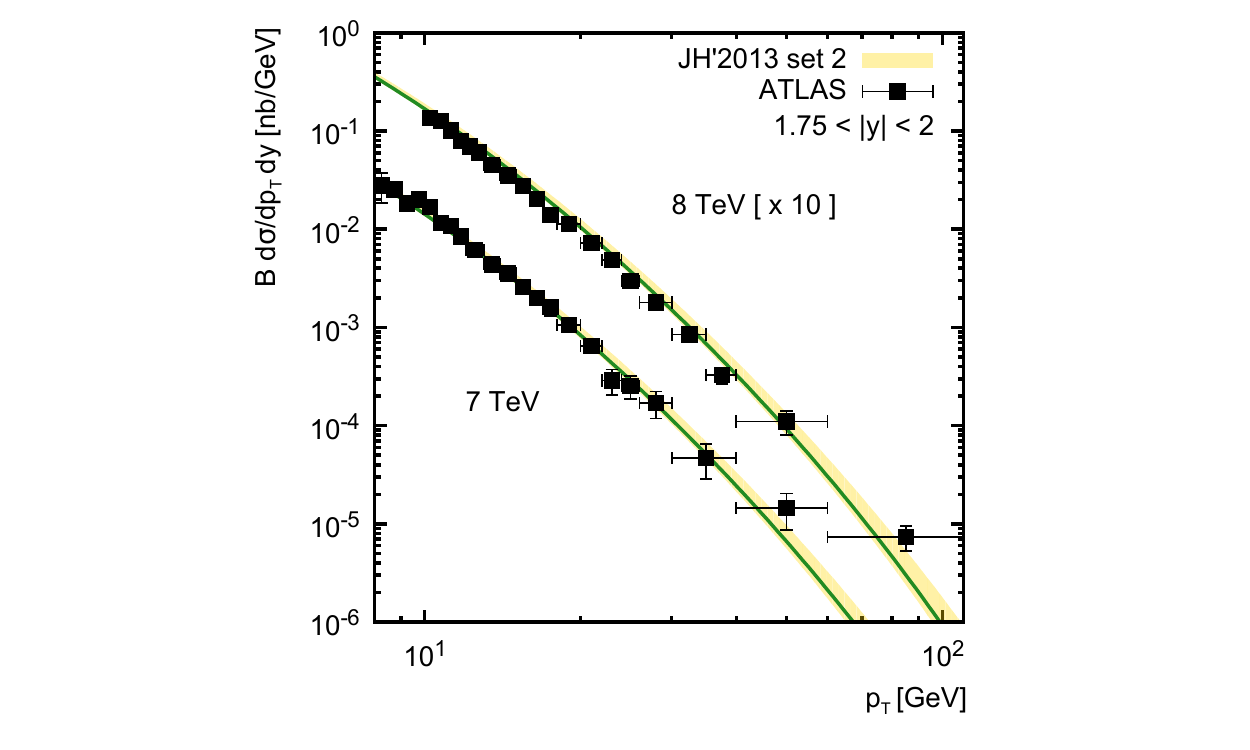}
\caption{The double differential cross sections of inclusive non-prompt $\psi(2S)$ meson
production at $\sqrt s = 7$ and $8$~TeV as function of $\psi(2S)$ transverse momentum.
Notation of curves is the same as in Fig.~1. The experimental data are from ATLAS\cite{2}.}
\label{fig4}
\end{center}
\end{figure}

\begin{figure}
\begin{center}
\includegraphics[width=7.9cm]{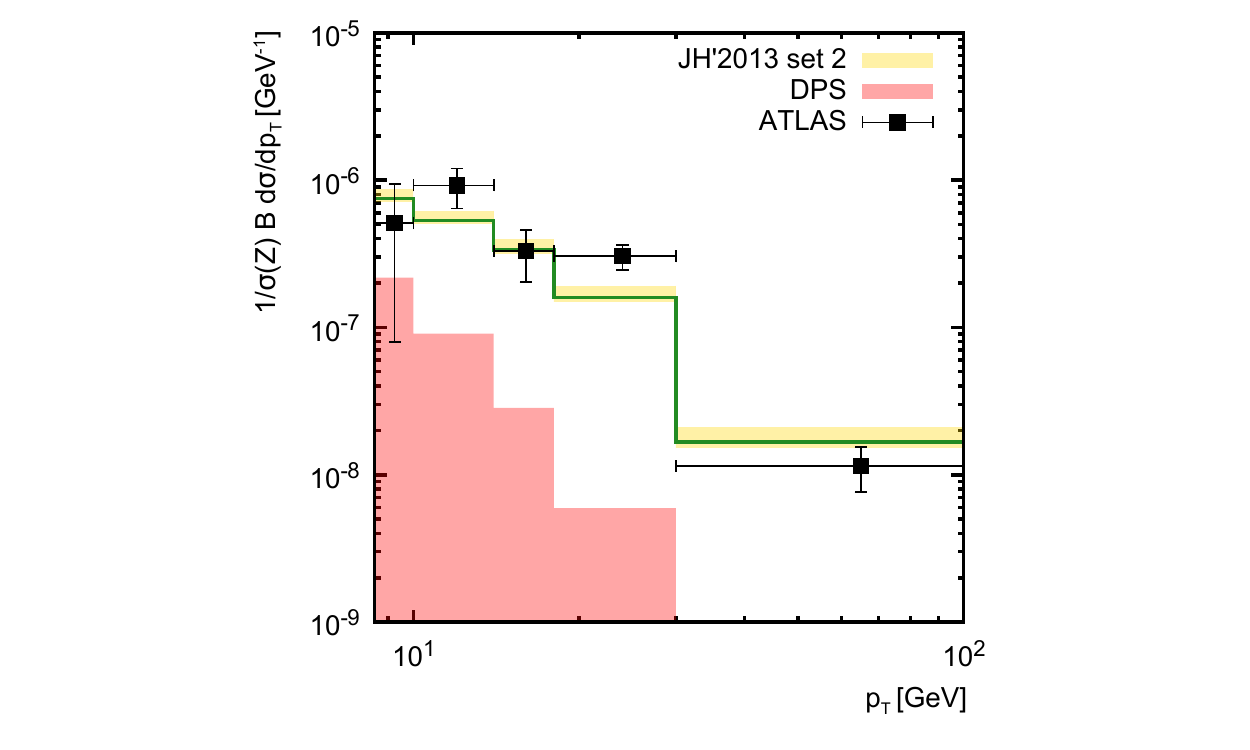}
\includegraphics[width=7.9cm]{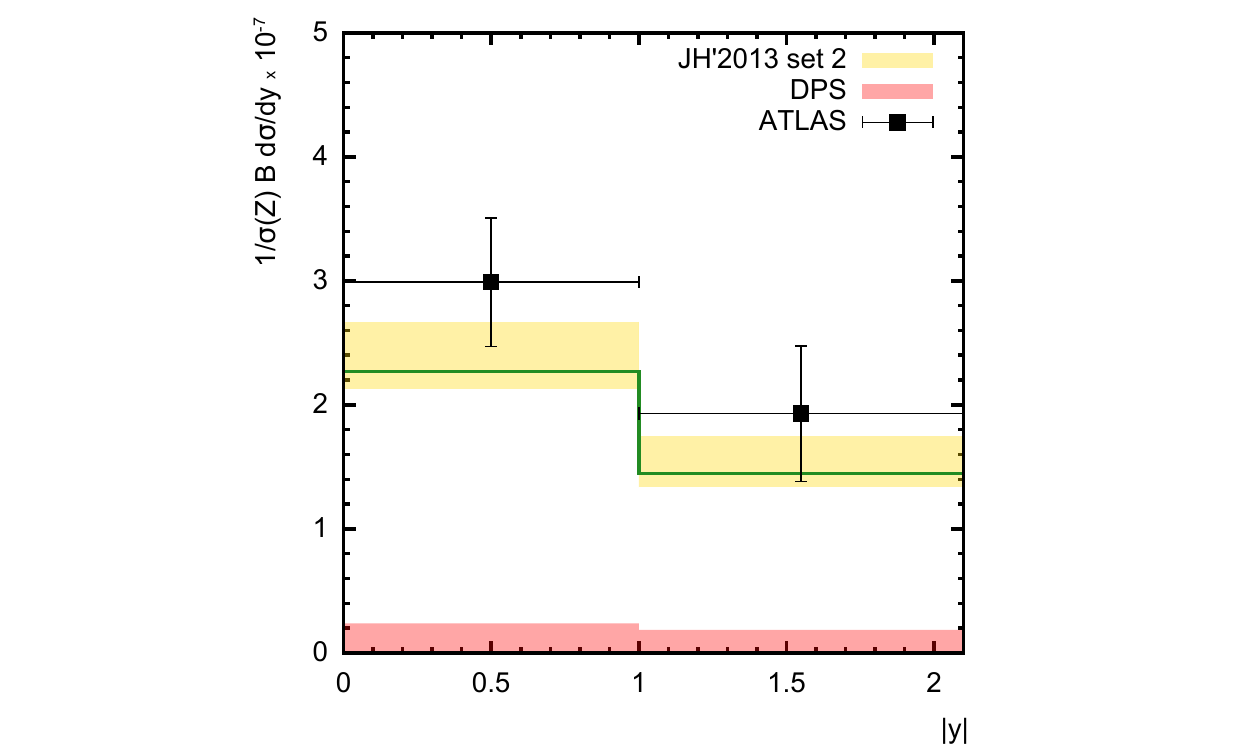}
\caption{The differential cross sections of associated non-prompt $J/\psi + Z$ production 
at $\sqrt s = 8$~TeV as functions of $J/\psi$ transverse momentum and rapidity.
Notation of curves is the same as in Fig.~1.
The estimated DPS contributions are shown separately and not summed with SPS ones.
 The experimental data are from ATLAS\cite{6}.}
\label{fig5}
\end{center}
\end{figure}

\end{document}